# `pracpac`: Practical R Packaging with Docker

**VP Nagraj**
Signature Science, LLC

**Stephen D. Turner**
Signature Science, LLC

March 21, 2023

## Abstract

R packages are the fundamental units of reproducible code in R, providing a mechanism for distributing user-developed code, documentation, and data. Docker is a virtualization technology that allows applications and their dependencies to be distributed and run reproducibly across platforms. The pracpac package provides an interface to create Docker images that contain custom R packages. The pracpac package leverages the renv package management tool to ensure reproducibility by building dependency packages inside the container image mirroring those installed on the developer's system. The pracpac package can be used to containerize any R package to deploy with other domain-specific non-R tools, Shiny applications, or entire data analysis pipelines. The pracpac package is available on CRAN (`https://cran.r-project.org/package=pracpac`), and source code is available under the MIT license on GitHub (`https://github.com/signaturescience/pracpac`).

## 1 Introduction

### 1.1 Background

R packages are the fundamental units of reproducible code in R (Wickham 2015). Docker is a virtualization technology that can be used to bundle an application and all its dependencies in a virtual container that can be distributed and deployed to run reproducibly on any Windows, Linux or MacOS operating system. Here we describe the **pracpac** package, which provides a templating library for building a Docker image containing an R package and other arbitrary domain-specific code and data. The pracpac package is designed for R package developers who want to ship a package and other ancillary functionality in a Docker container.

### 1.2 Related work

The pracpac package is conceptually inspired by usethis (Wickham, Bryan, and Barrett 2022), another templating package that automates and standardizes common package development tasks. The usethis package is called directly by the R package developer (e.g., `usethis::use_test("my_tests")`), and functions often write output directly to the active project (usually a package).

The renv package (Ushey 2022) is a tool to manage dependency versions of R packages. The package allows a user to initialize a project-local environment with a private R library, snapshot to save the state of the project library to a "lock file" that explicitly specifies all package versions and dependencies, and then restore the state of the project as specified in the lock file. This ensures that packages and their dependencies are frozen at their specific version at the time the project was initialized, allowing future analyses to pick up from the same R package versions frozen in the lock file. The pracpac package leverages renv to ensure that the environment used by the package developer (i.e., the versions of dependencies that are *known* to work) is mirrored in the Docker image to be built.

The Rocker project (Boettiger and Eddelbuettel 2017) provides pre-built Docker images containing R and optionally other tools such as RStudio, the Tidyverse (Wickham et al. 2019) suite of packages, LaTeX and other publishing tools, and geospatial analysis packages. The pracpac package defaults to using Rocker images as base images upon which to build.

The dockerfiler package (Fay et al. 2023) provides an interface for creating Dockerfiles (the de facto format for defining instructions to build images) from R. The rang package (Chan and Schoch 2023) has similar goals, but with the focus on reconstructing historical R computational environments which haven't been fully declared. The outsider package (Bennett et al. 2020) has a different focus, with the goal being to call Docker containers from within R. While these three tools operate at the intersection between Docker and R, they each solve different problems than pracpac.

### 1.3 Motivation

The pracpac package builds on functionality from some of the related work described above to deliver a novel tool for R package developers. Specifically, prapcac makes it trivial for developers to incorporate an R package into a Docker image. There are a number of use cases for the "R package + Docker" pattern, several of which are demonstrated later in this manuscript. Whatever the intended usage, pracpac significantly reduces the technical complexity of file manipulation (i.e., what files to create and where to save them in the directory structure) and reproducibility (i.e., ensuring that the package dependencies distributed in the container align with working versions).

## 2 The pracpac R package

The pracpac package is designed for developers who aim to ship their package and other arbitrary or domain-specific code as a Docker image. The pracpac package is strongly oriented towards reproducibility – by default, pracpac uses renv to ensure versions of all dependencies of an R package are frozen with the versions on the developer's system at the time of creation. An overview of the pracpac package workflow is shown in Figure 1. Below we describe the implementation of pracpac and demonstrate use cases.

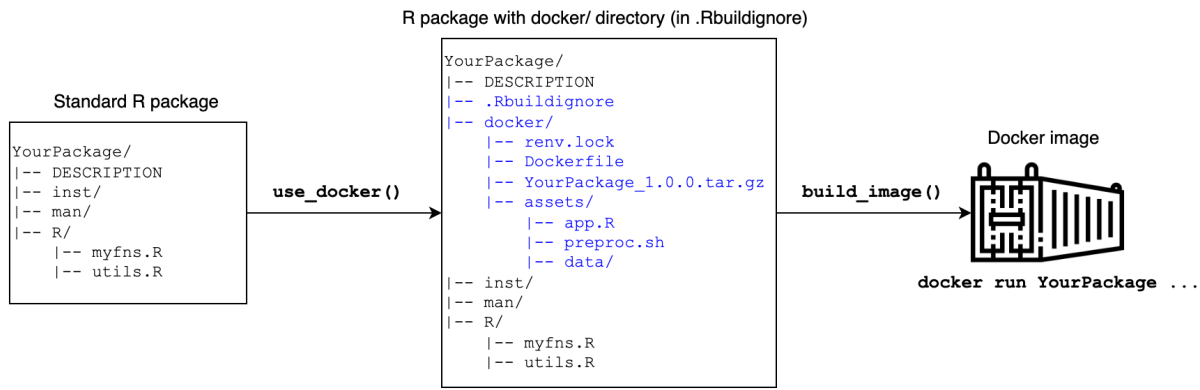

Figure 1: The pracpac workflow. The developer starts with a standard R package (left). When `use_docker` is run, a docker directory is created, the package source tar.gz is built and placed into that directory, a Dockerfile is created, and (depending on the use case) the optional asset placeholder(s) will be created as well. Using `build_image()` after any other optional editing of the Dockerfile will produce a Docker image with the R package, all its dependencies, and other assets as defined by the user.

### 2.1 Implementation

The pracpac package provides a wrapper function, `use_docker()`, which sequentially runs all the individual functions show in Table 1. These individual functions can be run independently as well. Using default settings, `use_docker()` will (1) create a Docker build directory (defaulting to `docker/` within the R package root) if one doesn't already exist, (2) evaluate what package dependencies are required by the package being built, determine which versions of these packages are on the user's system, and will create a `renv.lock` file to freeze these dependencies at the current version on the developer's system, (3) create a Dockerfile using a specified base image and optional template, (4) create placeholder assets such as arbitrary domain-specific code, shiny `app.R` files, etc., and place them into the Docker build directory, and (5) build the R package and copy it to the Docker build directory.

The `use_docker()` function includes an option to build the image based on the files in the Docker build directory. However, given that the developer will likely prefer to customize the Dockerfile and any assets prior to build, this option is set to `FALSE` by default. The developer may use the Docker API directly, a client in R (e.g., stevedore), or the pracpac `build_image()` function to build the image. If the pracpac method is used, by default the image will be tagged with the version in the R package `DESCRIPTION`.

Table 1: The main exported functions in the pracpac package.

| Function | What it does |
| --- | --- |
| `use_docker()` | Wrapper function that runs all of the pracpac templating functions with sensible defaults. |
| `pkg_info()` | Collects information about the package under development, such as the package name, version, dependencies, etc. This information is used by downstream functions. |
| `renv_deps()` | Creates a `renv.lock` file with all the current package's dependencies. The package developer is writing package code and tests using whichever version of the package's dependencies that are installed on the host system at the time of development, which may not necessarily be the most recent version of these dependent packages. `renv_deps()` determines the package's dependencies, then internally runs `renv::snapshot()` to create a lock file. This `renv.lock` file is placed inside the Docker build directory. |
| `add_dockerfile()` | Constructs a Dockerfile from a template. The developer optionally specifies a use case (e.g., `shiny`, `pipeline`, or `rstudio`). `add_dockerfile()` will start `FROM` a default image (e.g., `rocker/shiny:latest`, or `rocker/ver:latest`), but this can be overwritten to select a different base image (e.g., `alpine:3.17`). This function will create a `Dockerfile` with instructions to install the package's dependencies. |
| `add_assets()` | Adds example assets for a specified use case. For example, a shiny `app.R` for a Shiny app, or placeholder R and shell scripts for a pipeline use case. |
| `build_pkg()` | Builds the `package.tar.gz` source and copies it to the Docker build directory alongside the Dockerfile and other assets. |
| `build_image()` | Builds the image from the Dockerfile created by `add_dockerfile()`, tagging the image with the version pulled from the package's `DESCRIPTION` file using `pkg_info()`. |

### 2.2 Use cases

Below we provide several examples of use cases where the pracpac "R package + Docker" pattern may be useful. This is by no means exhaustive, and the pattern of delivering an R package within a Docker image may prove useful in other scenarios not described here as well. Please see the package vignettes for additional documentation and examples of use cases.

Note that in each example that follows, the resulting Dockerfile demonstrates `use_docker()` if it were run using the source for the pracpac package itself. In practice, the built R package copied and installed into the image would be the `tar.gz` for the developer's own package.

#### 2.2.1 Default

The `use_case="default"` option to `use_docker()` will create a minimal Dockerfile using a standard base image (`rocker/r-ver:latest` by default, but this can be changed with function arguments). The Dockerfile will include instructions to install package dependencies (with or without renv), and will install the developer's package into the image. No additional downstream steps (e.g., COPY, RUN, CMD, or ENTRYPOINT) are added to the resulting Dockerfile. The Dockerfile that follows was created with the "default" use case and renv option set to TRUE.

```
FROM rocker/r-ver:latest

## copy the renv.lock into the image
COPY renv.lock /renv.lock

## install renv and biocmanager
RUN Rscript -e 'install.packages(c("renv","BiocManager"), repos="https://cloud.r-project.org")'

## set the renv path var to the renv lib
ENV RENV_PATHS_LIBRARY renv/library

## restore packages from renv.lock
RUN Rscript -e 'renv::restore(lockfile = "/renv.lock", repos = NULL)'

## copy in built R package
COPY pracpac_0.1.0.tar.gz /pracpac_0.1.0.tar.gz

## run script to install built R package from source
RUN Rscript -e "install.packages('/pracpac_0.1.0.tar.gz', type='source', repos=NULL)"
```

### 2.2.2 Pipeline

The `use_case="pipeline"` option to `use_docker()` will create a Dockerfile with instructions to install package dependencies (with or without renv), and will install the developer's package into the image. This will also create a placeholder `assets/` directory which will be copied into the image. The `assets/` directory could contain scripts with optional preprocessing steps, domain-specific code executed inside the container, and/or postprocessing analysis, all coordinated by a "run" script. Additionally, the developer may wish to install additional system libraries or compiled code. The Dockerfile includes boilerplate and comments to motivate usage. The Dockerfile that follows was created with the "pipeline" use case and renv option set to TRUE.

```
FROM rocker/r-ver:latest

COPY renv.lock /renv.lock
RUN Rscript -e 'install.packages(c("renv","BiocManager"), repos="https://cloud.r-project.org")'
ENV RENV_PATHS_LIBRARY renv/library
RUN Rscript -e 'renv::restore(lockfile = "/renv.lock", repos = NULL)'

COPY pracpac_0.1.0.tar.gz /pracpac_0.1.0.tar.gz
RUN Rscript -e "install.packages('/pracpac_0.1.0.tar.gz', type='source', repos=NULL)"

# Placeholder for demonstration purposes
# Edit this Dockerfile for your specific needs
# Note that it may be necessary to install domain-specific tools
# yum update && yum install -y <pkg1> <pkg2> <pkg3>
# wget http://<url>/source.tar.gz && tar xzf source.tar.gz && cd source && make && make install

# Shows how to copy pipeline scripts from assets
# Note the following uses the files created during templating and copied to assets/
COPY assets/pre.R /pre.R
COPY assets/post.R /post.R
COPY assets/run.sh /post.sh

# Example runs a bash script wrapper
# Calls to run pre/post processing in R
CMD ["bash", "/run.sh"]
```

### 2.2.3 Shiny

The `use_case="shiny"` option to `use_docker()` will create a Dockerfile with instructions to install package dependencies (with or without renv), and will install the developer's package into the image. By default the base image for this use case will be `rocker/shiny:latest`. This will also create a placeholder `assets/app.R` which can be replaced by the developer's own `app.R`, which presumably uses the custom R package on which pracpac is being called. An alternative Shiny usage may be to distribute the app itself as a function in the custom R package on which pracpac is being called. In either case, the Shiny Server can be started with `docker run --rm -it -p 3838:3838 {image:tag}` (where `{image:tag}` is the image name and tag issued following Docker build). The Dockerfile that follows was created with the "shiny" use case and renv option set to `FALSE`.

```
FROM rocker/shiny:latest

COPY pracpac_0.1.0.tar.gz /pracpac_0.1.0.tar.gz

RUN Rscript -e 'install.packages("BiocManager")'
RUN Rscript -e "BiocManager::install(c('magrittr','glue','fs','rprojroot','renv','pkgbuild'), update=FALSE, ask=FALSE)"
RUN Rscript -e "install.packages('/pracpac_0.1.0.tar.gz', type='source', repos=NULL)"

COPY assets/app.R /srv/shiny-server

CMD ["/usr/bin/shiny-server"]
```

### 2.2.4 RStudio

The `use_case="rstudio"` option to `use_docker()` will create a Dockerfile with instructions to install package dependencies (with or without renv), and will install the developer's package into the image. By default the base image for this use case will be `rocker/rstudio:latest`. The resulting Dockerfile includes a call to launch the RStudio Server application in the `CMD` step. The RStudio Server can be started with `docker run --rm -it -p 8787:8787 {image:tag}` (where `{image:tag}` is the image name and tag issued following Docker build). The Dockerfile that follows was created with the "rstudio" use case and renv option set to `FALSE`.

```
FROM rocker/rstudio:latest

COPY pracpac_0.1.0.tar.gz /pracpac_0.1.0.tar.gz

RUN Rscript -e 'install.packages("BiocManager")'
RUN Rscript -e "BiocManager::install(c('magrittr','glue','fs','rprojroot','renv','pkgbuild'), update=FALSE, ask=FALSE)"
RUN Rscript -e "install.packages('/pracpac_0.1.0.tar.gz', type='source', repos=NULL)"

# Optionally expose RStudio Server on a different port
EXPOSE 8787
CMD ["/init"]
```

## 3 Conclusions

The pracpac package provides a developer-facing interface to create Docker images from within an R package development workflow. The pracpac package uses renv by default, bolstering reproducibility by ensuring packages in the container image are frozen at the version being used by the developer on the host system. We have used pracpac to build and deploy numerous containerized applications that include custom R package drivers and domain-specific data processing pipeline tools (including non-R source code and compiled executables).

The pracpac package is available on CRAN (https://cran.r-project.org/package=pracpac), and source code is available under the MIT license on GitHub (https://github.com/signaturescience/pracpac). Documentation and vignettes are available at https://signaturescience.github.io/pracpac/.

## 4 Acknowledgements

The authors have no competing financial interests to disclose. This research received no external funding, and was supported wholly by internal funding from Signature Science, LLC.

The authors would like to thank Chloé Skye Nagraj for creating pracpac's hex sticker artwork.